\documentclass[preprint,12pt]{elsarticle}
\usepackage{graphicx,color,mathrsfs,flafter,amsmath,amssymb,amsfonts}

\journal{Physics Letters B}

\begin{document}

\begin{frontmatter}

\title{Event-by-event fluctuations of magnetic and electric fields in heavy ion collisions}

\author[ab]{Adam Bzdak}
\address[ab]{RIKEN BNL Research Center, Brookhaven National Laboratory,\\Upton, NY 11973, USA}
\ead{abzdak@bnl.gov}

\author[vs]{Vladimir Skokov}
\address[vs]{Physics Department, Brookhaven National Laboratory, \\Upton, NY 11973, USA}
\ead{vskokov@bnl.gov}

\begin{abstract}
We show that fluctuating proton positions in the colliding nuclei generate, on the event-by-event basis, very strong magnetic and electric fields in 
the direction both parallel and perpendicular to the reaction plane.
The magnitude of $E$ and $B$ fields in each event is of the order of $m_{\pi }^{2}\approx 10^{18}$ Gauss. Implications on the observation of electric dipole in heavy ion collisions is discussed, and the possibility of measuring the electric conductivity of the hot medium is pointed out.
\end{abstract}


\date{\today}

\end{frontmatter}

\section{Introduction}

\begin{figure}[h]
\begin{center}
\includegraphics[scale=0.8]{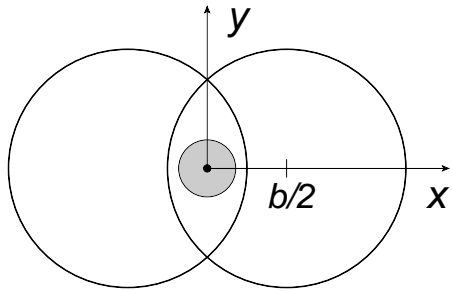}
\end{center}
\caption{The transverse plane of a peripheral heavy ion collision with the
impact parameter $b$. We calculate the magnetic $B$ and electric $E$ fields
at $t=0$ (time of the collision) and $\vec{x}=0$ (denoted by a black dote)
and we also average over a transverse area denoted by a grey circle. }
\label{fig_AA}
\end{figure}

The importance of the strong magnetic field created in heavy ion collisions
has been recently emphasized in Refs.~\cite%
{Kharzeev:2007jp,Kharzeev:2009fn,Kharzeev:2010gr}, where the chiral magnetic
effect has been proposed and studied. The chiral magnetic effect results in
the electric current along the direction of the magnetic field $\vec{B}$.
Consequently, the measurable charge separation may be observed in the
direction perpendicular to the reaction plane ($y$ axis in Fig.~\ref{fig_AA}%
) -- the dominant direction of the magnetic field averaged over events~\cite%
{Kharzeev:2007jp,Skokov:2009qp}. Recently, the STAR collaboration has
published~\cite{:2009txa} first results on the charge-dependent two-particle
correlations with respect to the reaction plane, which are qualitatively
consistent with the chiral magnetic effect expectations; this conclusion,
however, is under intensive debates (see, e.g., Refs.~\cite%
{Pratt:2010zn,Bzdak:2009fc,Muller:2010jd,Teaney:2010vd,Ma:2011um}%
).

As was mentioned earlier, according to early estimates (see Refs.~\cite%
{Kharzeev:2007jp,Skokov:2009qp}) the magnetic field perpendicular to the
reaction plane in heavy ion collisions is dominant and its magnitude may
reach very high values up to dozens of $m_{\pi }^{2}\approx 10^{18}$ Gauss.
However, in all previous analysis the fluctuation of the magnetic field in
each event was disregarded, and instead the averaged value over many events
was used to calculate $\langle \vec{B}\rangle $. The importance of the
event-by-event calculation becomes apparent after calculating the $x$
component of the magnetic field at $\vec{x}=0$ (denoted by a black dot in
Fig. \ref{fig_AA}).
Indeed, averaging over many events we obtain $\left\langle
B_{x}\right\rangle =0$ (as we explain later), suggesting that the $x$
component of the magnetic field can be neglected. In this Letter, we argue
that owing to fluctuating proton positions in both nuclei, $B_{x}$ becomes
comparable to $B_{y}$ on the event-by-event basis. We also calculate the
electric field $E$ and show that the $x$ and $y$ components of $B$ and $E$
fields are of the same order of magnitude, at least at the early stage of
the collision, when the hot medium response may be neglected.

In the next Section we describe our approach in detail and present our
results. Comments and summary are presented in Section $3$ and Section $4$,
respectively.

\section{Calculation}

The electric and magnetic fields at a position $\vec{x}$\ and observation
time $t$ can be calculated in each event according to the following
equations~\cite{Landau}%
\begin{eqnarray}
e\vec{E}(t,\vec{x}) &=&\alpha _{\mathrm{EM}}\sum\nolimits_{n}\frac{%
1-v_{n}^{2}}{R_{n}^{3}\left( 1-[\vec{R}_{n}\times \vec{v}_{n}]^{2}/R_{n}^{2}%
\right) ^{3/2}}\vec{R}_{n},  \notag \\
e\vec{B}(t,\vec{x}) &=&\alpha _{\mathrm{EM}}\sum\nolimits_{n}\frac{%
1-v_{n}^{2}}{R_{n}^{3}\left( 1-[\vec{R}_{n}\times \vec{v}_{n}]^{2}/R_{n}^{2}%
\right) ^{3/2}}\vec{v}_{n}\times \vec{R}_{n},  \label{EB}
\end{eqnarray}%
where sums are performed over all protons in both nuclei. The fine-structure
constant $\alpha _{\mathrm{EM}} = e^{2}/4\pi \approx 1/137$ and $\vec{R_{n}}=%
\vec{x}-\vec{x}_{n}(t)$, where $\vec{x}_{n}$ is a position of proton moving
with the velocity $\vec{v}_{n}$. If $\vec{v}_{n}$ is non-zero only in the $z$
direction, then $(\vec{R}_{n}\times \vec{v}_{n})^{2}=R_{n,\perp
}^{2}v_{n,z}^{2}$, where the transverse vector between a proton and the
observation point $\vec{R}_{n,\perp }=\vec{x}_{\perp }-\vec{x}_{n,\perp }$
does not depend on time $t$.\footnote{%
It should be noted that the retardation effects are already taken into
account and $\vec{R_{n}}$ depends explicitly on the observation time $t$.}

In our calculation we sample proton positions $\vec{x}_{n}$ at $t=0$
according to the Woods-Saxon distribution with the standard parameters \cite%
{Alver:2008aq}. We also assume that both nuclei are infinitely thin; and
that all target and projectile protons move with the same velocity $\vec{v}%
_{n}^{\,\text{targ}}=[0,0,v_{z}]$ and $\vec{v}_{n}^{\,\text{proj}%
}=[0,0,-v_{z}]$, respectively. The value of $v_{z}$ is defined by the
collision energy $\sqrt{s}$ and the proton mass $m_{p}$, $v_{z}^{2}=1-\left(
2m_{p}/\sqrt{s}\right) ^{2}$. Finally, when calculating $\vec{B}$ and $\vec{E%
}$ at a given point $\vec{x}$ we take into account only those protons which
are not closer than $r_{\text{cut}}=0.3$ fm to the observation point $\vec{x}
$. This cut-off is to be implemented owing to the singularities of Eqs.~(\ref%
{EB}) at $R_{n}\rightarrow 0$. The value $r_{\text{cut}}=0.3$ fm was fixed
as an effective distance between partons in a nucleon. By a variation of $r_{%
\text{cut}}$ in the range from $0.3$ fm to $0.6$ fm we affirm weak cut-off
dependence of the fields.

\begin{figure}[h]
\begin{center}
\includegraphics[scale=0.5]{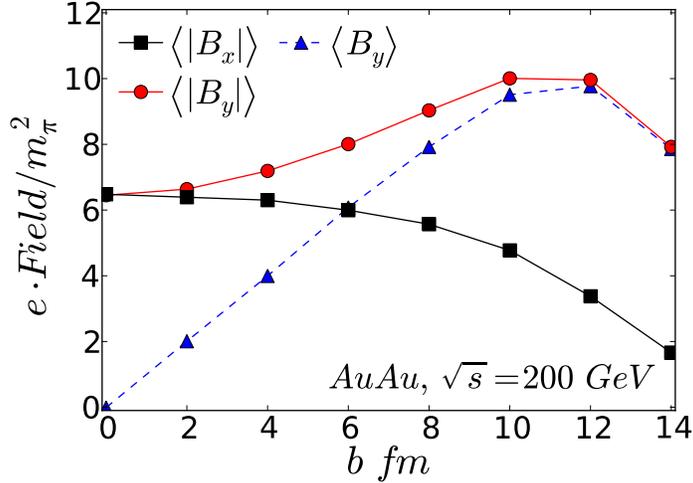}
\end{center}
\caption{The mean absolute value of the magnetic field at $t=0$ and $\vec{x}%
=0$ as a function of impact parameter $b$ for $AuAu$ collision at $\protect%
\sqrt{s}=200$ GeV. Fluctuation of proton positions lead to non-zero values
of $\left\langle |B_{x}|\right\rangle $ that are comparable to $\left\langle
|B_{y}|\right\rangle $.}
\label{fig_B}
\end{figure}

We performed our calculations at $t=0$ and at $\vec{x}=0$ (denoted by a
black dot in Fig. \ref{fig_AA}). The results for magnetic field in $AuAu$
collisions at $\sqrt{s}=200$ GeV are presented in Fig. \ref{fig_B}.

As seen from Eq. (\ref{EB}) $\left\langle B_{x}\right\rangle =0$ at $\vec{x}%
=0$.\footnote{%
Indeed, one nucleus gives $\left\langle B_{x}\right\rangle \sim $ $%
\left\langle y_{n}\right\rangle =0$, in contrast to $\left\langle
B_{y}\right\rangle \sim $ $\left\langle x_{n}\right\rangle \neq 0$ if $b>0$,
see Fig. \ref{fig_AA}.} However, in each event the $x$ component of the
magnetic field can be huge and vanishes only after averaging over events. To
study the magnitude of the magnetic field disregarding its direction from
one event to another we consider the average (over events) absolute value%
\footnote{%
First we calculate, e.g., $B_{x}$ in an event from all protons and after
that we take the absolute value. Next we calculate average over events.} of
the magnetic and electric fields, $\left\langle |B_{x,y}|\right\rangle $ and 
$\left\langle |E_{x,y}|\right\rangle $. Due to fluctuations of the proton
positions we obtain comparable numbers for $\left\langle
|B_{x}|\right\rangle $ and $\left\langle |B_{y}|\right\rangle $ suggesting
that on the event-by-event basis we should expect huge fields both in $x$
and $y$ directions. Since the chiral magnetic effect leads to the electric
current along the magnetic field, our result indicate that, in principle,
the chiral magnetic effect may take place not only in the $y$ direction but
also in the $x$ direction.

\begin{figure}[h]
\begin{center}
\includegraphics[scale=0.5]{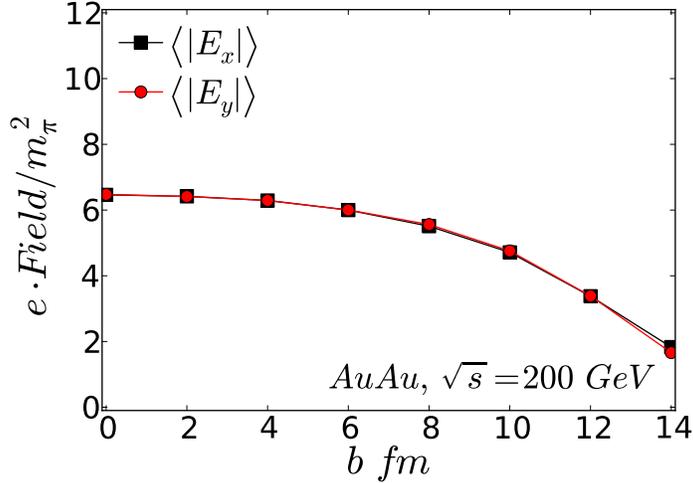}
\end{center}
\caption{The mean absolute value of the electric field at $t=0$ and $\vec{x}%
=0$ as a function of impact parameter $b$ for $AuAu$ collision at $\protect%
\sqrt{s}=200$ GeV. Fluctuation of proton positions lead to non-zero values
of $\left\langle |E_{x}|\right\rangle \approx \left\langle
|E_{y}|\right\rangle $.}
\label{fig_E}
\end{figure}

The results for the electric field are shown in Fig. \ref{fig_E}. 

The symmetry of the system presented in Fig. \ref{fig_AA} implies that at $%
\vec{x}=0$ the average value of the electric field $\left\langle
E_{x}\right\rangle =\left\langle E_{y}\right\rangle =0$. However, as seen in
Fig. \ref{fig_E}, fluctuations lead to $\left\langle |E_{x}|\right\rangle
\approx \left\langle |E_{y}|\right\rangle $ with magnitude of the order of $%
m_{\pi }^{2}$. It is interesting to notice that $x$ and $y$ components of
the electric field are almost identical to the $x$ component of the magnetic
field%
\begin{equation}
\left\langle |B_{x}|\right\rangle \approx \left\langle |E_{x}|\right\rangle
\approx \left\langle |E_{y}|\right\rangle .  \label{eq}
\end{equation}%
This conclusion can be drawn easily from the analysis of Eqs. (\ref{EB}).
The fields $B_{x}$ and $E_{x,y}$ are driven only by fluctuations of the
proton positions and since $v_{z}\approx 1$ it is evident that Eq. (\ref{eq}%
) is valid. For peripheral collisions, the $y$ component of the magnetic
field is influenced not only by the fluctuations, but also (mainly) by the
geometry of the collision, as seen from Fig. \ref{fig_B}, where $%
\left\langle |B_{y}|\right\rangle $ and $\left\langle B_{y}\right\rangle $
are compared. Thus we expect $\left\langle |B_{y}|\right\rangle $ to be
larger than the fields in Eq.~(\ref{eq}).

\begin{figure}[h]
\begin{center}
\includegraphics[scale=0.5]{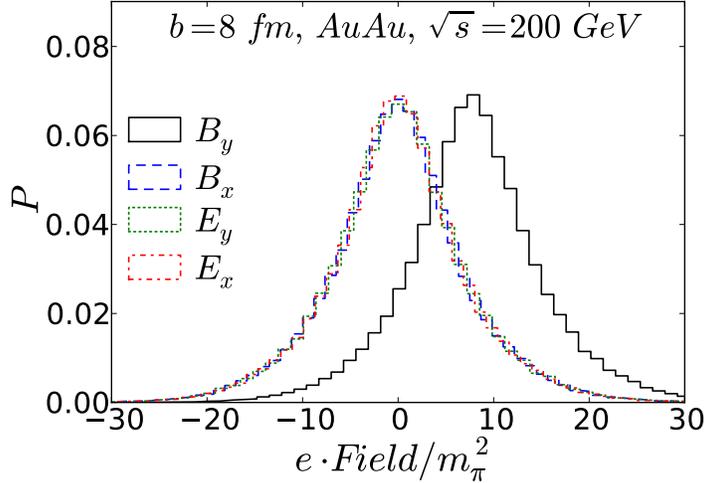}
\end{center}
\caption{Normalized event-by-event histograms of $x$ and $y$ components of
magnetic $B$ and electric $E$ fields at $t=0$ and $\vec{x}=0$ at the impact
parameter $b=8$ fm. As expected for peripheral collisions, $B_{y}$
distribution is shifted away from zero. $B_{x},$ $E_{x}$ and $E_{y}$
histograms are practically indistinguishable.}
\label{fig_hist}
\end{figure}

To provide more complete information on the fluctuating values of $B$ and $E$
fields, we show in Fig. \ref{fig_hist} the event-by-event histograms for $%
B_{x,y}$ and $E_{x,y}$ at the impact parameter $b=8$ fm. As expected $B_{y}$
distribution is shifted away from zero and $B_{x},$ $E_{x}$ and $E_{y}$
distributions are practically indistinguishable, consistent with Eq. (\ref%
{eq}). 
We checked that for $b=0$ histograms look very similar, the only difference
is $B_{y}$ that is centered around zero being indistinguishable from $B_{x},$
$E_{x}$ and $E_{y}$.

\section{Comments}

Several comments are in order.

(i) We performed analogous calculations for $CuCu$ collisions at the top
RHIC energy and for $PbPb$ collisions at the LHC energy. We found very
similar qualitative behavior - in fact at $t=0$ and $\vec{x}=0$ the
following approximate scaling holds for $\left\langle |B_{x,y}|\right\rangle 
$, $\left\langle |E_{x,y}|\right\rangle $%
\begin{equation}
\frac{\text{Field}}{m_{\pi }^{2}}\propto \frac{\sqrt{s}}{m_{p}}f(b/R_{A}),
\end{equation}%
where $R_{A}$ is the radius of the appropriate nucleus, $\sqrt{s}$ is the
center of mass energy and $m_{p}$ is a proton mass. In the first approximation, the
function  $f(x)$ is universal.% and depends weakly on $\sqrt{s}$ and $b$.

\begin{figure}[h]
\begin{center}
\includegraphics[scale=0.5]{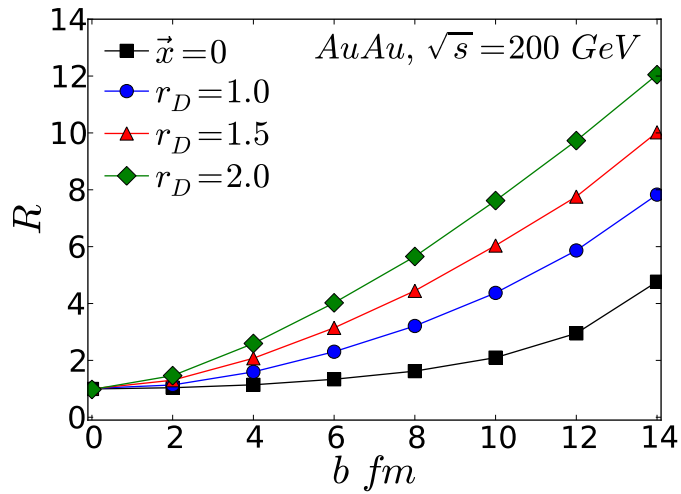}
\end{center}
\caption{The ratio $R=\left\langle |B_{y,\Omega }|\right\rangle
/\left\langle |B_{x,\Omega }|\right\rangle $ as a function of the impact
parameter. $B_{x}$ and $B_{y}$ are integrated over the transverse domain $%
\Omega $ of radius $r_{D}$ [fm]. For comparison, we also show the ratio at $%
\vec{x}=0$.}
\label{fig_r}
\end{figure}

(ii) To check the uniformity of the magnetic and electric fields in each
event, we integrated $E$ and $B$ over the transverse spherical domain $%
\Omega $ (denoted in Fig. \ref{fig_AA} by a grey circle), i.e., 
\begin{equation}
\left\langle |B_{x,\Omega }|\right\rangle =\left\langle \left|
\int\nolimits_{\Omega }B_{x}(\vec{x}_{\perp })d^{2}\vec{x}_{\perp }\right|
\right\rangle /(\pi r_{D}^{2}),
\end{equation}%
and analogously for $B_{y},$ $E_{x}$ and $E_{y}$. Here $r_{D}$ is the radius
of the domain which we varied from $r_{D}=1$ fm to $r_{D}=2$ fm. As seen
from Fig. \ref{fig_r}, for $b=4$ fm the ratio $R=\left\langle |B_{y,\Omega
}|\right\rangle /\left\langle |B_{x,\Omega }|\right\rangle $ is changing
from $1.6$ to $2.5$ (with increasing $r_{D}$), as compared to $1.14$ at $%
\vec{x}=0$. For $b=8$ fm $R$ is changing from $3.2$ to $5.6$, as compared to 
$1.6$ at $\vec{x}=0$. %These results suggest that indeed $\left\langle
%|B_{x}|\right\rangle $ and $\left\langle |B_{y}|\right\rangle $ are of the
%same order of magnitude. 
The ratio $\left\langle |E_{y,\Omega
}|\right\rangle /\left\langle |E_{x,\Omega }|\right\rangle $ is always very
close to $1$, and again the $x$ and $y$ components of $E$ are almost
identical to the $x$ component of $B$. 

As seen in Fig.~\ref{fig_r}, for small b the ratio $R$ weakly depends on the size
of the domain in contrast to large b, where the $r_D$ dependence is
significant. We performed our calculation up to $r_D = 2$ fm (size of
the domain equals 4 fm), which in peripheral collisions covers almost
the whole interaction region. In practical applications to the chiral magnetic effect,
the fluctuations of the magnetic field are important only in the region with a characteristic size
corresponding to the size of a sphaleron. 
At least at early stage of a collision, where the influence of the magnetic field is significant,
this size is expected to be of the order of 1 fm (the characteristic size of a sphaleron is less 
than the magnetic screening length, which is $(\alpha_s T)^{-1}$  in the weak coupling limit). This
suggests, as seen in Fig. 5, that, indeed,  $\left\langle
|B_{x}|\right\rangle $ and  $\left\langle |B_{y}|\right\rangle $ are of the same order and should be treated equally in calculations  of the chiral magnetic effect. 

(iii) In an actual experiment, the reconstructed reaction plane (the
so-called event plane) is, in general, tilted with respect to the one
defined by the geometry of colliding nuclei. Thus a possible
experimental measurement that is sensitive to the direction of the
magnetic (electric) field in each event, will reveal the components
$B_x'$ ($E_x'$) and $B_y'$ ($E_y'$), that are linear combinations of
the original ones $B_x$ ($E_x$) and $B_y$ ($E_y$). This mixing
provides an additional reinforcement to our conclusion that the
different field components are comparable in magnitude.

(iv) Since the electric conductivity is significantly larger than the one
associated with the magnetic field (at least an order of magnitude, see
Refs.~\cite{Ding:2010ga,Yamamoto:2011ks}) it is quite clear that the
current created by the electric field $j_{E}$ dominates over the one due to
the chiral magnetic effect $j_{B}$. As seen from Fig. \ref{fig_B} and Fig. %
\ref{fig_E} we expect the ratio $\langle|\vec{j}_{E}|\rangle /\langle |\vec{j%
}_{B}|\rangle $ to be maximal for central collisions and minimal (but still
larger than $1$) for very peripheral collisions.

(v) Our results suggest that the strong electric current should not pose a
significant problem for the correlation observable $\left\langle \cos (\phi
_{1}+\phi _{2}-2\Psi _{\text{RP}})\right\rangle $ proposed in Ref. \cite%
{Voloshin:2004vk} to measure the chiral magnetic effect. Indeed, this
observable is sensitive to the difference between correlations in-plain and
out-of-plain and, in the first approximation, $j_{E}$ seems to be the same
in $x$ and $y$ directions, in contrast to $j_{B}$ created in peripheral
collisions (but not in central collision, where we expect $j_{B,x}\approx
j_{B,y}$). This point, however, should be investigated in more elaborated
studies, that take into account realistic temperature- and density-
dependence of the electric conductivity.

(vi) In principle, by measuring the charge separation\footnote{%
For instance, in the way proposed in Ref. \cite{Liao:2010nv}.} in central
collisions one may perform rough estimate of the electric conductivity $%
\sigma $ of the matter created in heavy ion collisions. Assuming the
dominance of the electric current in charge separation process in central
collisions, and estimating effective lifetime $t_{\mathrm{eff}}$ and spatial
extensions of the hot spot, where the electric field is almost constant and
homogeneous $E_{\mathrm{eff}}$, we obtain $\sigma \sim \frac{Q}{t_{\mathrm{%
eff}}S_{\perp }E_{\mathrm{eff}}}$. Here $Q$ is the dipole charge and $%
S_{\perp }$ is the area perpendicular to the dipole axis. This problem is
currently under our investigation.

(vii) Our estimates of the magnetic and electric fields are valid only at the
early stage of the collision. At later stages, the magnetic response from
the created medium becomes increasingly important \cite%
{Tuchin:2010vs,Landau8}, and may lead to substantial increase of the
magnetic field over the electric one. For quantitative analysis of this
effect, more elaborated studies are required~\cite{VK}.

\section{Summary}

In this Letter we argue that owing to the fluctuating proton positions in
the relativistic heavy ion collisions, the strength of the magnetic and
electric fields are comparable (both in the direction parallel and
perpendicular to the reaction plane) in the event-by-event analysis. Thus,
for the observables, that are sensitive to the electric and magnetic fields,
it is essential to take into account event-by-event field fluctuations.

Our result in combination with the lattice QCD calculations of the electric
and the chiral magnetic conductivities suggest, that the possible charge
separation measured separately in-plane and out-of-plane will be dominated
by the electric current, which in principle allows to measure the electric
conductivity of the hot medium created in heavy ion collisions. However,
since the electric field is not correlated with reaction plane, the
difference between correlations in-plain and out-of-plain in peripheral
collisions should be insensitive to the electric field.

\section*{Acknowledgments}

We thank Dmitri Kharzeev and Larry McLerran for helpful discussions. This
investigation was supported by the U.S. Department of Energy under Contract
No. DE-AC02-98CH10886 and by the grant N N202 125437 of the Polish Ministry
of Science and Higher Education (2009-2012).

\end{document}